# Gravitational Collapse and Radiation of Grand Unified Theory

*Yi-Fang Chang*
*Department of Physics, Yunnan University, Kunming, 650091, China*
(E-mail: yifangchang1030@hotmail.com)

**Abstract** The infinite gravitational collapse of any supermassive stars should pass through an energy scale of the grand unified theory (GUT). After nucleon-decays, the supermassive star will convert nearly all its mass into energy, and produce the radiation of GUT. It may probably explain some ultrahigh energy puzzles in astrophysics, for example, quasars and gamma-ray bursts (GRB), etc. This is similar with a process of the Big Bang Universe with a time-reversal evolution in much smaller space scale and mass scale. In this process the star seems be a true white hole.

For supermassive stars the theory indicates that there is nothing that can stop the cores of these heavyweight stars from collapsing indefinitely [1]. Usually it is a certainty that a black hole will fall finally into a singularity, although some gravitational theories, for instance, quantum gravity and superstring theory, may solve the problem of singularity, and the quantum effects can prevent a star from collapsing to a singularity [2]. Sonego, et al., defined optical geometry for spherically symmetric gravitational collapse, which provides a natural physical interpretation for derivations of the Hawking effect based on the "moving mirror analogy" [3]. Recently, Whiting discussed identifying the singular field for self-force evaluation [4]. Mino reviewed theoretical progress in the self-force problem of a general relativistic two-body system, and discussed the radiation reaction formula [5]. Mitsuda, et al., analyzed the stability of self-similar solutions for perfect fluid gravitational collapse [6].

The grand unified theory (GUT) is introduced in order to unify strong, electromagnetic and weak interactions in particle physics [7]. For example, in Coleman-Weinberg SO(10) GUT [8],

$$SO(10) \xrightarrow{M_X} G \xrightarrow{M_R} SU(3)_c \otimes SU(2)_L \otimes U(1)_Y, \qquad (1)$$

where $M_X$ is of order $10^{15} - 10^{16} GeV$ to be compatible with the lower limit on proton decay, and G is one of the intermediate symmetry groups, and the scale $M_R$ is the one relevant for neutrino physics. We proposed that GUT of particle physics could be applied to the supermassive stars, whose infinite gravitational collapse will pass through an energy scale of GUT, so some ultrahigh energy puzzles in astrophysics may be explained [9].

In astrophysics it is very successful that GUT is used to determine the symmetry breaking in the very early universe. A characteristic energy scale of the simplest SU(5) GUT model is of order $10^{15}$ GeV. It corresponds the initial $10^{-36}$ second in the Big Bang Universe. Now GUT comes to a standstill, and is difficult in the application of astronomy because of ultrahigh energy and proton-decays. Weinberg [10] has proposed that the rates of baryon-nonconserving processes, like proton decay, are very small at ordinary energies. But if the slowness of these processes is due to the large mass of intermediate vector of scalar "Xboson"$\pm$ which mediate baryon nonconservation, then at very high temperatures with $kT \cong M_X$, the baryon-nonconserving processes would have rates comparable with those of other processes. Weinberg discussed this possibility for $kT \geq M_X$. We analyze these cases, and propose that these difficulties may be solved for the infinite gravitational collapse.

A gravitational collapsing process from a supermassive star to black hole and singularity is very similar with a process of the Big Bang Universe. Only the former is a time-reversal evolution with in much smaller space scale and mass scale. Both end points are the singularities with an infinite density, and the two processes pass through various energy scales. For black holes, the Planck scale $10^{19}$ GeV and the Planck density $10^{94} gcm^{-3}$ have been discussed, which will pass through a GUT's energy scale necessarily. We think a collapsing process from star to black hole (esp. without the event horizon) and singularity may possibly repeat many processes of the Big Bang Universe.

In a collapsing process the potential energy of stars will become large continuously. If a supermassive star collapses to a singularity, its potential energy will be infinite, which is independent of the concrete model. So long as GUT appears in an evolutionary process of infinite gravitational collapse for any supermassive star, protons and neutrons will decay

$$p \to e^+ + \pi^0 ....... n \to \nu + \pi^0 ....... \qquad (2)$$

Detailed calculations show that the decay-branches are

$$p \to \begin{matrix} e^+\pi^0 & e^+\omega & e^+\rho & e^+\eta \\ 30\% & 30\% & 14\% & 4\% \end{matrix} ........ \qquad (3)$$

Further, $\pi^0 \to \gamma\gamma (98.8\%), \gamma e^+ e^- (1.2\%),$ (4)

$$\omega \to \pi^+\pi^-\pi^0 (89.6\%), \gamma e^+ e^- (8.7\%)...... \qquad (5)$$

Such, a supermassive star will emit the radiation of GUT, and convert mass into huge energy with efficiency up to $(m_p - m_e)/m_p$ =1834/1835=99.95% for the energy scale. For the existence of the process, the most important clue is that conditions of nucleon-decay must be satisfied. It is also independent of GUT's concrete model or energy scale.

Moreover, the energy scale of other GUT's models, except SU(5) model, may be different. For example, supersymmetric SO(10) GUTs have the intermediate scale $10^{10} - 10^{12}$ GeV [11-13], $E_6$ model has $10^4 - 10^6$ GeV [14,15], and general supersymmetric models with the B-nonconserving processes could proceed among $10^{16}$ GeV and 100 GeV depending upon the concrete model.

The Kerr-Newman spacetime has the line element:

$$ds^2 = (1 - \frac{2mr - Q^2}{\rho^2})dt^2 - \frac{\rho^2}{\Delta}dr^2 - \rho^2 d\theta^2 - [(r^2 + a^2) \\ + \frac{(2mr - Q^2)a^2 \sin^2\theta}{\rho^2}]\sin^2\theta d\varphi^2 + \frac{2(2mr - Q^2)a\sin^2\theta}{\rho^2}dtd\varphi. \qquad (6)$$

where $\Delta = r^2 + a^2 + Q^2 - 2mr$, $\rho^2 = r^2 + a^2 \cos^2\theta$. The Kerr-Newman black hole is the most general stationary black hole with parameters M(mass), J=M$a$(angular momentum) and Q(charge). Its radius is

$$R_\pm = \frac{GM}{c^2} \pm \frac{1}{c^2}\sqrt{(GM)^2 - GQ^2 - (\frac{cJ}{M})^2} . \qquad (7)$$

It is one of Schwarzschild black hole $R = 2GM/c^2$ in a case that Q=J=0. On the other hand, when

$$(GM)^2 - GQ^2 - (cJ/M)^2 < 0, \qquad (8)$$

i.e., $Q^2 + \sqrt{Q^4 + (2cJ)^2} > 2GM^2 > Q^2 - \sqrt{Q^4 + (2cJ)^2}$, (9)

this has not an event horizon. Thus an external observer can see all processes, in which a black hole collapses to a singularity with mass and charges.

The condition (9) may hold completely. For instance, it becomes

$$Q^2 > GM^2 > 0, \qquad (10)$$

for a black hole with charge Q and without rotation, i.e., J=0. It is namely a static electric force larger than a gravitational force of the same distance. Or it becomes $J > GM^2/c$ for Q=0.

The infinite gravitational collapse should possess following characteristics: 1. Emission photons and neutrinos, etc., with ultrahigh energy, because

$$\pi^0 \to \gamma\gamma, E(\gamma) = (140/2)MeV = 70MeV, \quad (11)$$

$$\gamma\gamma \to e^+e^- \to \text{any hadrons}, \ e^+e^- \to \gamma\gamma \to \mu^+\mu^- \ldots\ldots \quad (12)$$

2. Disappearance of stars with 100% or near 100%. These processes (11) and (12) are going on again, then the photon energies decrease continually. These rays connect probably with two notable puzzles in astrophysics [1,16]: quasars and gamma-ray bursts (GRB). Moreover, this process may also correspond to some sources of ultrahigh energy cosmic rays.

Many quasars are radio sources, which come from compact clouds of electrons moving with approximate light speed. The models of quasars include colliding stars, quasar superstars, giant pulsars and black hole theory, etc. The present theories suppose that the quasars are created in early stages of evolutionary universe. The GRBs are the sudden flashes of the hard X/gamma-rays in the sky. It was discovered that the spatial distribution of GRB's sources is isotropic [17], and require both enormous total energy and the concentration of that energy into a small mass. The sources emitted energies during a few seconds or a few minutes larger than that emitted by Sun during $10^{10}$ years, and disappeared mystically. The energies emit for GB971214 and GB990123 even up to $3\times 10^{53} erg$ and $3.4\times 10^{54} erg \approx 1.9 M_O$. This implies that the source must be a compact object, perhaps accreting black hole of stellar size. The GRB's photon energy may be 170KeV-1GeV [18-21].

Total energy of Sun is $E = mc^2 = 2\times 10^{33} \times 9\times 10^{20} gcm/\sec = 1.8\times 10^{54} erg$. The known power of quasars is about $10^{47} erg/\sec$, so continuously emitting energy of quasars per year($= 3.1536\times 10^7$ sec) is $3.1536\times 10^{54} erg$, which consumes only about $2M_O$. If we take that the energy put out by Sun in its whole life (10 billion years) is $10^{51} erg$ [1], the mass of quasars will be of the order $1000 M_O$. While the quasar superstar model may include $10^6 - 10^8$ solar masses [1], in some other models even as much as $10^{15} M_O$ [22,23].

So far, some hundred models of GRB and quasars have been proposed, but none is confirmed fully. Some energy-sources of quasars convert mass of quasars into kinetic energy, with nearly 100% efficiency [24]. A mechanism for the conversion is gravitational collapse [25], but the calculations of such high efficiency are very difficult [26,27]. Usually a standard model of the compact radio cores causes a rotating supermassive ($10^8 - 10^{10} M_O$) black hole and accretion disk [28]. Astronomers proposed the neutron star accreting model, the shock emission model [29], the neutron starquake as a potential local model [30] and very rapid accretion into a black hole model. At present, the usually adopted models are mainly the neutron star-neutron star merging model and the black hole-neutron star merging model [31]. Another important type of model for GRB is the collapsing star, i.e., Collapsar, or Hypernova [32-34], and a neutron star collides with a black hole or two neutron stars coalesce, etc. But, a probability of these collisions should be very small in the Universe, although it will increase much in the core of Galaxy.

We develop a gravitational collapsing theory, and propose that quasars are probably the radiation source of GUT, whose differences from quasars produced by black holes are ultrahigh energy and terminal into almost nothing. It agrees with following characteristics of quasars [1]:

a. Ultrahigh energy sources. Quasars can outshine the entire Milky Way by a factor of 1000. While nearly all mass of this star may transform into energy.

b. High-speed electrons and photons may produce large redshifts, caused either by a local explosion, or from powerful gravitational fields within the objects themselves [24]. Then photon energy decreases, $E(\gamma) = 14 MeV$ for a redshift z=4.

c. The $\pi^0 \to \gamma\gamma$ produce two symmetric gamma-ray sources, which like "two engines mode"；± Other decays of nucleons will also form some particular directions of emitted energies.

d. Intermittent bursts of quasars correspond to continual collapse with some layers, which is analogous to the structure of neutron star [35].

e. The powerhouse is a small object.

f. Rapid motion is associated with the production of energy.

For GRB, Band, et al., studied the time-averaged GRB spectrum, and proposed the formula [36]

$$N_E(E) = A(\frac{E}{100keV})^\alpha \exp(-\frac{E}{E_0}), \quad (\alpha-\beta)E_0 \geq E \text{ at low energy}, \quad (13)$$

$$N_E(E) = A'(\frac{E}{100keV})^\beta, \quad\quad\quad (\alpha-\beta)E_0 < E \text{ at high energy}. \quad (14)$$

Here $\alpha \approx -1 \sim -0.5, \beta \approx -2 \sim -3, E_0 \approx 0.1 \sim 1 MeV$ [37,38].

A character of GRB is the initial detection eight hours after the main burst, and faded afterglows within a few days with a power-law decay function [39]. When $\alpha \approx -1, \beta \approx -2$, the formula (13) and (14) may be a unified form

$$N_E(E) = C(\frac{E}{100keV})^\alpha \frac{1}{e^{E/E_0}-1}. \quad (15)$$

It is namely the Planck black radiation formula and the Bose-Einstein distribution.

Further, the GRB spectrum is consistent with some formulas of particle physics. For instance, the multiplicity distribution which associated with the geometric bremsstrahlung model is the Gamma Distribution [40]

$$D_\Gamma = \frac{\beta^\alpha}{\Gamma(\alpha)} x^{\alpha-1} \exp(-\beta x). \quad (16)$$

In the generalized bremsstrahlung model of high energy collision, the multiplicity distribution function is [41]:

$$W_n(s,t) = \frac{1}{n!}[b(s,t)]^n \exp[-b(s,t)]. \quad (17)$$

The early universe possesses ultrahigh energy, many quantities, for example, the number of species of particles, the density of energy, as well as the asymptotic solution of Einstein's equation, etc., all accord with Eq.(16) [42]. The total scattering cross section of the fireball model at high energy is also [43]:

$$\sigma(S) = \beta^2(0)(S/S_0)^{\alpha(0)-1}. \quad (18)$$

The quantitative comparisons of these formulas show that the radiations of GUT in the infinite gravitational collapses are possibly the sources of GRB. It may explain some GRB's characteristics, in particular, with enormous total energy and mystical disappearance. Of course, some black holes may probably introduce quasars and GRB, etc. Moreover, GUT and black hole, etc., in high energy astrophysics energy violate possibly the Pauli exclusion principle. In 1984, based on some experiments and theories of particles at high energy, I suggested that particles at high energy possess a new statistics unifying Bose-Einstein(BE) and Fermi-Dirac(FD) statistics, so PEP would not hold at high energy [44-46].

Finally, it is interesting to compare with the white hole [47], which was first suggested in 1964 by I.Novikov and M.Hjellming as a possible source of quasar energy. White hole is a time-reversed black hole, but it is a mathematical creature [1]. The process of infinite gravitational collapse possesses some properties of white holes, in which baryon-number is nonconservation. Perhaps, this is a true white hole. The infinite gravitational collapses pass through the energy scale of GUT, which may be a concrete astronomical mechanism on the creative cosmos [48]. In this case, the matter is released continually from supermassive stars, which is similar to little Universe on big bang. Such both are symmetric.